\documentclass[fleqn,twoside]{article}
\usepackage{espcrc2}
\usepackage{graphicx}

\title{Hardware and software status of QCDOC\,\thanks{Based on talks
    by K.~Petrov and T.~Wettig at Lattice 2003, Tsukuba.}}

\author{%
P.A.~Boyle\address[Ed]{Department of Physics and Astronomy, University 
  of Edinburgh, Edinburgh EH9 3JZ, Scotland}\address[Columbia]{Department 
  of Physics, Columbia University, New York, NY 10027, USA},
D.~Chen\address[IBM]{IBM T.J.~Watson Research Center, Yorktown
  Heights, NY 10598, USA}, 
N.H.~Christ\addressmark[Columbia],
M.~Clark\addressmark[Ed],
S.D.~Cohen\addressmark[Columbia], 
C.~Cristian\addressmark[Columbia], 
Z.~Dong\addressmark[Columbia],
A.~Gara\addressmark[IBM],
B.~Jo\'o\addressmark[Ed],
C.~Jung\addressmark[Columbia]\address[BNL]{Physics Department, Brookhaven
  National Laboratory, Upton, NY 11973, USA},
C.~Kim\addressmark[Columbia], 
L.~Levkova\addressmark[Columbia], 
X.~Liao\addressmark[Columbia],
G.~Liu\addressmark[Columbia],
R.D.~Mawhinney\addressmark[Columbia],
S.~Ohta\address{Institute for Particle and Nuclear Studies,
  KEK, Tsukuba, Ibaraki, 305-0801, Japan}\address[RBRC]{RIKEN-BNL 
  Research Center, Brookhaven National Laboratory, Upton, NY, 11973, USA},
K.~Petrov\addressmark[Columbia]\addressmark[BNL],
T.~Wettig\addressmark[RBRC]\address{Department of Physics, Yale
  University, New Haven, CT, 06520, USA},
A.~Yamaguchi\addressmark[Columbia]}
       
\begin{document}

\begin{abstract}
  QCDOC is a massively parallel supercomputer whose processing nodes
  are based on an application-specific integrated circuit (ASIC).
  This ASIC was custom-designed so that crucial lattice QCD kernels
  achieve an overall sustained performance of 50\% on machines with
  several 10,000 nodes.  This strong scalability, together with
  low power consumption and a price/performance ratio of \$1 per
  sustained MFlops, enable QCDOC to attack the most demanding lattice
  QCD problems.  The first ASICs became available in June of 2003, and
  the testing performed so far has shown all systems functioning
  according to specification.  We review the 
  hardware and software status of QCDOC and present performance
  figures obtained in real hardware as well as in simulation.
  \vspace{1pc}
\end{abstract}

\maketitle

\section{Introduction}

Continued advances in commodity processing and networking hardware
make PC (or workstation) clusters a very attractive alternative for
lattice QCD calculations \cite{lippert}.  Indeed, there are quite a
few important problems that can be addressed on PC clusters, and many
lattice physicists are taking advantage of this opportunity.  However,
for the most demanding problems in lattice QCD, e.g.\ dynamical
fermion simulations with realistic quark masses, one would like to
distribute the global volume over as many nodes as possible, resulting
in a very small local volume per node.  PC clusters are inadequate to
deal with this case because the communications latency inherent in
their networking hardware implies that the local volume must not be
chosen too small if a reasonable sustained performance is to be
achieved.  In other words, for typical lattice QCD problems PC
clusters do not scale well beyond a few hundred nodes.

In custom-designed supercomputers such as QCDOC \cite{qcdoc,chep03} and
apeNEXT \cite{ape}, the communications hardware is designed to reduce
the latencies and to assist critical operations (such as global sums)
in hardware.  As a result, these machines are significantly more
scalable and allow for much smaller local volumes.  In addition, they
provide low power consumption, a small footprint, and a very low
price/performance ratio per sustained MFlops.  On the downside, the
development effort is considerably higher than for PC clusters, but
this effort is amortized by the unique strengths of these machines.

\begin{figure*}[!t]
  \vspace*{-3mm}\hspace*{-37mm}\includegraphics[width=195mm]{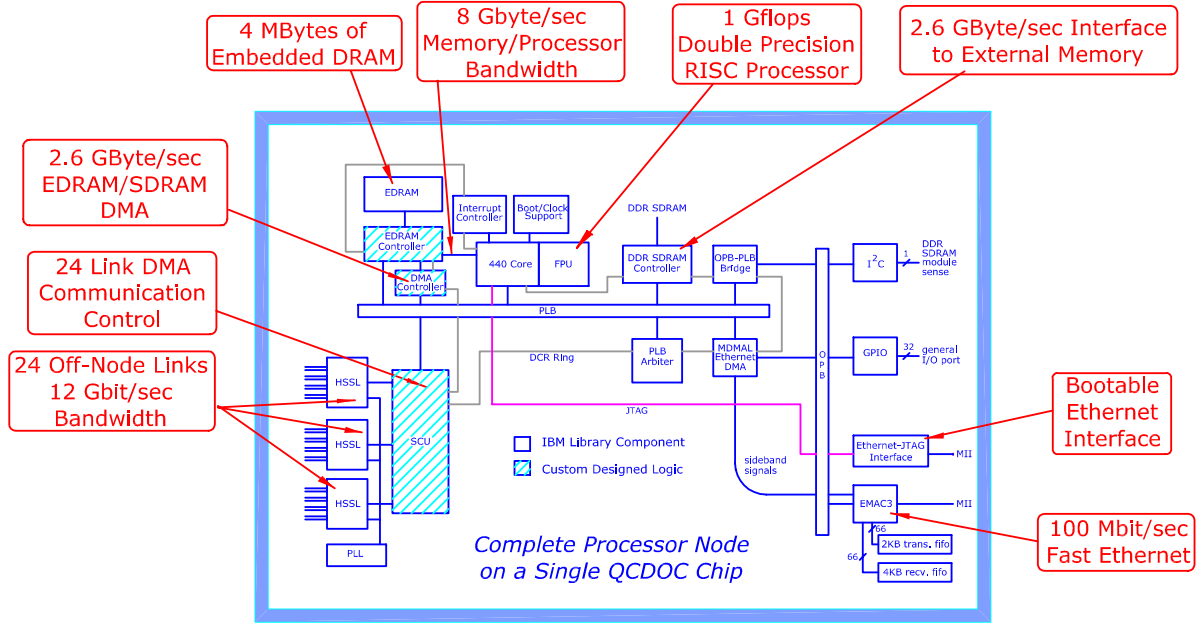}
  \vspace*{-30mm}
  \caption{Block diagram of the QCDOC ASIC, where the cross-hatched
    blocks have been custom-designed.}
  \label{fig:asic}
\end{figure*}


\section{Hardware}

The QCDOC hardware has been described in detail in several previous
publications, see Refs.~\cite{qcdoc,chep03}, therefore we only summarize its
most important features here.

\begin{figure}[!b]
  \centering
  \vspace*{-5mm}
  \includegraphics[width=\columnwidth]{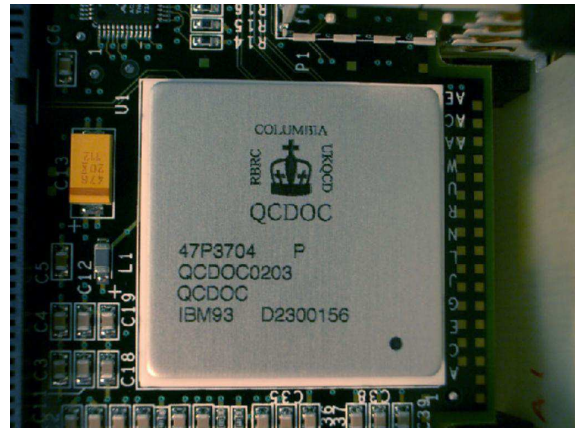}
  \vspace*{-10mm}
  \caption{Close-up view of a QCDOC ASIC.}
  \label{fig:asic_closeup}
\end{figure}

The QCDOC ASIC, shown schematically in Fig.~\ref{fig:asic}, was
developed in collaboration with IBM research and manufactured by IBM.
It contains a standard PowerPC 440 core running at 500 MHz, a 64-bit,
1 GFlops FPU, 4 MBytes of embedded memory (eDRAM), and a serial
communications interface (SCU) which has been tailored to the
particular requirements of lattice QCD.  The SCU provides direct
memory access, single-bit error detection with automatic resend, and a
low-latency pass-through mode for global sums.  Also on the chip are
several bus systems, controllers for embedded and external (DDR)
memory, an Ethernet controller, a bootable Ethernet-JTAG interface,
and several auxiliary devices (interrupt controller, I$^2$C interface,
etc.).  A picture of one of the first ASICs, delivered in June of
2003, is shown in Fig.~\ref{fig:asic_closeup}.

\begin{figure}
  \centering
  \includegraphics[width=\columnwidth]{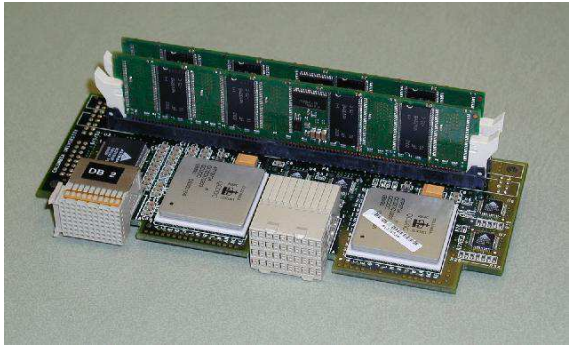}
  \vspace*{-10mm}
  \caption{A daughterboard with two QCDOC ASICs and two DDR DIMMs.}
  \vspace*{-3mm}
  \label{fig:db}
\end{figure}

The physical design of a large machine is as follows.  Two ASICs are
mounted on a daughterboard, together with two standard DDR memory
modules (one per ASIC) with a capacity of up to 2 GBytes each.  The
only other nontrivial components on the daughterboard, apart from a
few LEDs, are four physical layer chips for the MII interfaces (two
per ASIC) and a 4:1 Ethernet repeater which provides a single 100
Mbit/s Ethernet connection off the daughterboard.  A picture of the
very first two-node daughterboard is shown in Fig.~\ref{fig:db}.
\begin{figure}[!b]
  \centering
  \vspace*{-5mm}
  \includegraphics[width=\columnwidth]{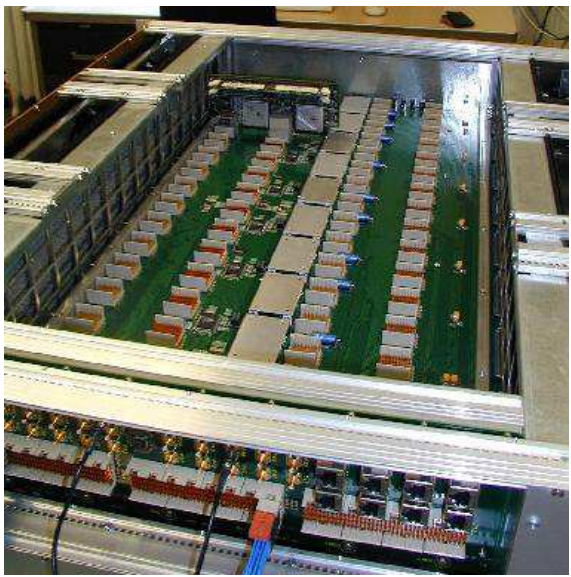}
  \vspace*{-10mm}
  \caption{A QCDOC motherboard populated by a single daughterboard.}
  \label{fig:mb}
\end{figure}
A motherboards holds 32 such daughterboards, eight motherboards are
mounted in a crate, and a large machine is built from the desired
number of crates.  A picture of a QCDOC motherboard is shown in
Fig.~\ref{fig:mb}.

The physics communications network of QCDOC is a 6-dimensional
torus with nearest-neighbor connections.  The two extra dimensions
allow for machine partitioning in software so that recabling is not
required.  A 64-node motherboard has a $2^6$ topology, with three open
dimensions and three dimensions closed on the motherboard (one of
which is closed on the daughterboard).  The SCU links run at 500
Mbit/s and provide separate send and receive interfaces to the forward
and backward neighbors in each dimension, resulting in a total
bandwidth of 12 Gbit/s per ASIC (of which 8 Gbit/s will be utilized in a
4-dimensional physics calculation).

\begin{figure}[!b]
  \centering
  \vspace*{-7mm}
  \includegraphics[width=\columnwidth]{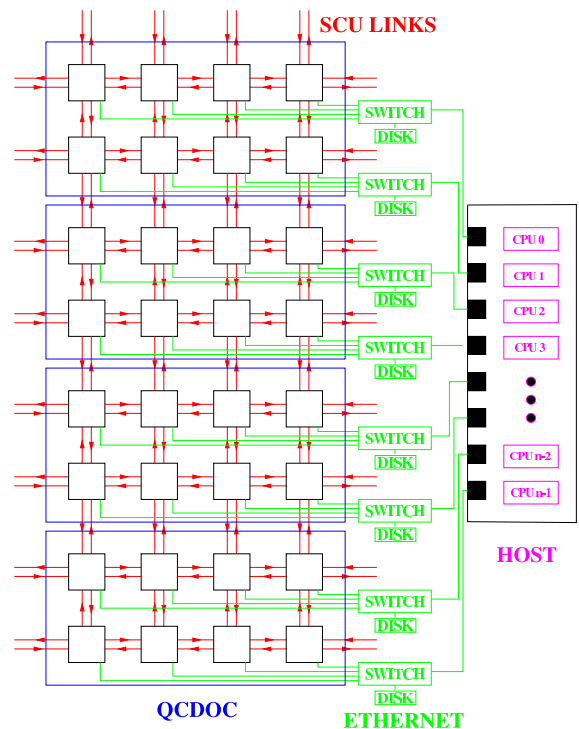}
  \vspace*{-10mm}
  \caption{A two-dimensional version of the network connections in a
    QCDOC machine.  Each small square corresponds to a processing node,
    and the large rectangles represent a motherboard (actual
    motherboards have 64 nodes).}
  \label{fig:network}
\end{figure}
In addition to the physics network, there is an Ethernet based network
for booting, I/O, and debugging, as well as a global tree network for
three independent interrupts.  The Ethernet traffic from/to each
motherboard proceeds at 800 Mbit/s to a commercial Gbit-Ethernet
switch tree, a parallel disk system, and the host machine.  The latter
will be a standard Unix SMP machine with multiple Gbit-Ethernet cards.
See Fig.~\ref{fig:network}.

As of the writing of this article (September 2003), all major
subsystems of the QCDOC ASIC have been tested in single-daughterboard
configurations (2 ASICs per daughterboard) using a temporary
test-jig.  This configuration allows non-trivial communications
between the two ASICs in one of the six dimensions; for the remaining
dimensions the ASIC communicates with itself in a loop-back mode.
Extensive memory tests with different sized commodity external DDR
SDRAM modules have been done, tests of the 4 MByte on-chip EDRAM
have been performed, and all the DMA units have been used.
High-performance Dirac kernels have been run for Wilson and ASQTAD
fermion formulations, confirming the performance figures given in
Table~\ref{tab:performance} and Ref.~\cite{chep03}.  No problems
with the ASIC have been found to date. 

With QCDOC motherboards now in hand for use (Fig.\ \ref{fig:mb}),
tests of 64 and 128 node machines are imminent.  In our test-jig
tests, the ASICs appear to be working well close to the target
frequency of 500 MHz.  With a fully populated motherboard, and the
more stable electrical environment provided by the motherboard as
compared to our simple test-jigs, we will soon be able to test a
large number of ASICs at the 500 MHz design frequency.  From
preliminary measurements, the power consumption per node is
about 5 W.

\section{Software}

One of the major goals of the development team was to make QCDOC
accessible and deployable to a large scientific community, so every
attempt has been made to allow users to use standard software tools
and techniques.  In order to reliably boot, monitor and diagnose the
10,000+ nodes in a large QCDOC machine, and to allow user
code maximum access to the high-performance capabilities of the
hardware, we are writing a custom operating system for QCDOC.
This is modeled on the QCDSP operating system, which
has successfully allowed the use of a machine with a similar number
of processing nodes.  The QCDOC operating system (QOS) has been
improved in myriad ways, continuing to focus on ease-of-use for
the end-user.

\subsection{Operating System}

We now give a more detailed list of the requirements of the operating
system and discuss how these features have been implemented.  In
addition to booting the machine, the QOS must diagnose hardware
errors and monitor the hardware during user program execution.
Monitoring is accomplished by regular inspection of on-chip registers
which monitor memory and communication link status, along with
interrupt handlers which are invoked by changes in hardware state
deemed crucial.  After booting, partitions of the machine may be
requested for interactive use by a user or the queuing system.
Each partition will only run a single user program at a time; all
multitasking will be handled by the host front-end.  During program
execution, the operating system provides host-QCDOC input/output
access using commodity Ethernet connections,
access to special QCDOC features such as
the mesh communication network and on-chip memory and access from
QCDOC processing nodes to a parallel disk array.  The QOS is designed
to coordinate the interactions between the host and QCDOC, insulating
the user from many QCDOC-specific details while still providing
detailed control over their execution environment.

Our solution is QOS, written largely from scratch in C++, C and
some assembler.  The software runs on the host computer and the QCDOC
nodes and consists of several parts, namely:
\begin{itemize}\itemsep 0mm
  \item [i] The Qdaemon on the host, which manages and monitors the
    entire QCDOC machine.  All interactions go through the Qdaemon.
  \item [ii] The Qclient layer on the host, which provides access to
    the Qdaemon for a variety of planned user interfaces, such as
    the batch system and Web applications, as well as the existing
    command line interface, the Qcsh.
  \item [iii] The Qcsh on the host, a modified version of tcsh, which
    is currently in use and allows complete control and use of QCDOC. 
  \item [iv]  The boot/run kernels on each node.
\end{itemize}

The Qdaemon is the heart of the operating system, and it is designed
and implemented using modern C++ techniques, e.g.\ it is heavily
templated, it employs POSIX threads to allow efficient use of a
multi-processor front end, its queue calls and object lists are
thread-safe, and it can drive multiple physical network interfaces.
The single Qdaemon on the host uses reliable UDP to transport
packets between QCDOC and the host, starting with basic UDP packets
in the boot procedure and adding an RPC protocol to this when the
run kernels are loaded.  The Qdaemon controls the IP addressing
scheme to the nodes, does the partitioning that users request,
monitors machine status, directs I/O to the host and evokes hardware
testing when needed.  The Qdaemon can be accessed by a ``root" user
for system management and monitoring, but most users will only have
access to a limited set of Qdaemon's capabilities, those appropriate
for the partition where they are running a job.  The Qdaemon 
also provides flexible software partitioning of QCDOC.

The Qclient is a library of interfaces which communicates with the
Qdaemon and can accept input from a variety of planned user tools.
To access the Qclient, and hence the Qdaemon, access to the host
computer must be gained by conventional authentication tools such
as ssh or a secure Web connection.  Currently, a command-line
interface, the Qcsh, communicates with the Qdaemon via the Qclient
library.

Following the successful QCDSP model, Qcsh
is a command-line interface.  Starting from a standard unix
shell, tcsh, extra built-in commands to control QCDOC have been
added.  Users can use normal shell redirection to control I/O from
QCDOC, programs can be executed on QCDOC and backgrounded in Qcsh
allowing further interaction between the Qcsh and Qdaemon (but not
QCDOC directly since it is executing a program), and standard shell
scripts can be used.  The special QCDOC commands in the Qcsh
all begin with the letter q.  Some
examples are: {\tt qinit} - establish a connection between the Qcsh and
the Qdaemon; {\tt qpartition\_create} - create a partition of QCDOC;
{\tt qboot}
- boot a QCDOC partition; {\tt qrun} - run a user program; {\tt qhelp}
- display available Qcommands.

The Qdaemon will only permit connections
from the local host, which allows a Unix socket to be passed to
the Qdaemon from a user interface like the Qcsh.  The Qdaemon then
does user I/O from/to this Unix socket and can determine Unix user
and group identities directly from the host OS kernel.  This is a
major simplification.  While the Qdaemon will be handling many
users' access to QCDOC, it does not have to open/close files and
concern itself with 
protections.  
This is handled by the initial user interface, thereby ensuring
correct ownership and permissions.

The first software loaded to a QCDOC node is the boot kernel, which
is loaded directly to the data and instruction caches of the 
440 core via the Ethernet-JTAG hardware.  Only the 440 core must
be functional for the boot kernel to execute.  When execution
begins, tests of the on-chip and DDR memory can be done.
Once the memory is known
to be working, the boot kernel enables the standard 100 Mbit Ethernet
port on the QCDOC ASIC and the run kernel is loaded down.  The run
kernel 
handles the activation of machine-global features such as the
nearest-neighbor communications network.  When these steps complete,
which should be on the scale of ten minutes even for a large machine,
it is available for general use.  All of the steps outlined above are
implemented and have been extensively used
on our first ASICs.

The run kernel provides support for users, including access to QCDOC
features via system calls.  Part of the strategy for using RPC for the
host--run-kernel communication is to avoid having a separate
communications protocol for disk access.  
We have written an NFS client for the run kernel that uses the standard
RPC-based NFS protocol. The client supports two mount points and
open/read/write/close functionality. This NFS support has already been
tested on QCDOC.
While providing a
standard user environment for high-performance 
computing,
the run-kernel will not support multi-tasking.  We have chosen to keep
the run-kernel lean and compact to ensure reliability and to keep the
software task bounded.

\subsection{Communications Software}

A major issue for a massively parallel computer is the effectiveness
of its message passing.  In hardware, QCDOC is a nearest-neighbor
mesh, the majority of QCD based communications are also nearest-neighbor
and consequently QOS natively supports nearest-neighbor communications
calls.  The QOS calls are implemented so as not to interfere with
the low latency of the QCDOC communications hardware.  The QCDOC
hardware 
supports efficient global sums (done via our pass-through
hardware), which are also accessible
via QOS calls.  We have also implemented the SciDAC Lattice QCD
Software Committee's QCD message passing (QMP) protocol on QCDOC.
This protocol supports nearest-neighbor communications, which we
efficiently map to the native QOS communications calls, as well as
arbitrary communications.  For the latter, we will
implement a Manhattan-style routing.

\subsection{Existing Software and Benchmarks}

A simple (but not trivial) consequence of QCDOC using a standard
processor from IBM's PowerPC line is the availability of a whole
arsenal of Open Source and commercial software tools, most notably
the GNU tools and IBM's tools.  On QCDOC, users will be able to
use the GNU toolset, the closest thing to a standard across computer
platforms. 
Additionally, we have access
to the high-performance commercial tools from IBM, such as the xlc/xlC
compilers which we have seen outperform the GNU compilers on PowerPC
platforms by as much as a factor of two.  We expect that users may do
initial code development with the GNU toolset and only compile with the
more restrictively-licensed IBM compilers when final performance is an
issue.

An issue for the larger potential user group for QCDOC is performance
of existing physics codes, not written for a QCDOC or QCDSP type
of computer architecture.  The MILC code was an obvious choice to
test, since it is one of the major lattice simulation codes and is
written in C.  The MILC code has been run on QCDOC, both the
simulator and now the actual ASIC, concentrating on the ASQTAD
action.  Unmodified MILC code gives performance in the few percent
range for small lattice volumes, which was easily improved by a
few standard C-code modifications as described in \cite{chep03}.  
A summary of the performance for single precision MILC code and
our double precision assembly code is given in Table
\ref{tab:performance}.

\begin{table}
\begin{tabular} {l|c|c|c} 
 Action & Vol.& Assem. & MILC \\ \hline
Wilson & $2^4$ & 47\%  \\
       & $4^4$ & 54\%  \\
\hline
Clover& $2^4$ & 56\% \\
      & $4^4$ & 59\% \\
\hline
Staggered & $2^4$ & 36\% & 17\% \\
          & $4^4$ &   & 21\% \\
Asqtad  &  $4^4 $ &  43\%  &  15\% \\
AF  &  $2^4$ & & 14\%\\
    &  $4^4$ & & 20\%\\

\end{tabular}
\caption{Performance for double precision assembly code and single
precision MILC code for various local lattice volumes and actions.
AF is the ASQTAD force term for the Hybrid Monte Carlo.}
\vspace*{-5mm}
\label{tab:performance}
\end{table}

\section{Schedule}

A 128-node prototype machine is currently being assembled at
Columbia.  Assuming that no major problems are found, two machines
of 10 TFlops (peak) each will be available to UKQCD and the RIKEN-BNL
Research Center by the summer of 2004.  The operating system and user
software needed to utilize these machines is progressing in hand
with the hardware developments.  Approval is pending for a 3 TFlops
(peak) machine at Columbia and a $\ge$ 20 TFlops (peak) installation
at BNL for the U.S. lattice community.

\bigskip

\noindent {\bf Acknowledgments.}
This work was supported in part by the U.S. Department of Energy, the
Institute of Physical and Chemical Research (RIKEN) of Japan, and the
U.K. Particle Physics and Astronomy Research Council.

\end{document}